\documentclass{ws-procs9x6}

\begin{document}

\begin{center}
Accepted for publication in the Proceedings of the ICATPP Conference on
Cosmic Rays for Particle and Astroparticle Physics,\\ Villa  Olmo (Como, Italy), 7--8 October, 2010, \\to be published by World Scientific (Singapore).
\end{center}
\vspace{-1.7cm}

\title{PROTON MODULATION IN THE HELIOSPHERE \\
FOR DIFFERENT SOLAR CONDITIONS \\
AND PREDICTION FOR AMS-02}

\author{P. Bobik$^{1}$, G. Boella$^{2,3}$, M.J. Boschini$^{2,4}$,
C. Consolandi$^{2}$, S. Della Torre$^{2,5}$, \\
M. Gervasi$^{2,3}$, D. Grandi$^{2}$, K. Kudela$^{1}$, E.
Memola$^{2}$, S. Pensotti$^{2,3}$, \\
P.G. Rancoita$^{2}$ and M. Tacconi$^{2}$}

\address{$^{1}$ Institute of Experimental Physics, Kosice (Slovak Republic) \\
$^{2}$Istituto Nazionale di Fisica Nucleare, INFN Milano-Bicocca, Milano (Italy) \\
$^{3}$Department of Physics, University of Milano Bicocca, Milano (Italy) \\
$^{4}$CILEA, Segrate (MI) (Italy) \\
$^{5}$Department of Physics and Maths, University of Insubria, Como (Italy)\\
E-mail: Massimo.Gervasi@mib.infn.it}

\begin{abstract}
Spectra of Galactic Cosmic Rays (GCRs) measured at the Earth are
the combination of several processes: sources production and
acceleration, propagation in the interstellar medium and
propagation in the heliosphere. Inside the solar cavity the
intensity of GCRs is reduced due to the solar modulation, the
interaction which they have with the interplanetary medium. We
realized a 2D stochastic simulation of solar modulation to
reproduce CR spectra at the Earth, and evaluated the importance in
our results of the Local Interstellar Spectrum (LIS) model and its
agreement with data at high energy. We show a good agreement
between our model and the data taken by AMS-01 and BESS
experiments during periods with different solar activity
conditions. Furthermore we made a prediction for the differential
intensity which will be measured by AMS-02 experiment.
\end{abstract}


\section{Introduction}

Models of the Heliosphere need to be very accurate in order to
reproduce the complex structure of the solar cavity and its effect
on the Cosmic Rays propagation. In particular simulations of CR
intensity have to be compared with experimental values, which have
been measured in the last years by several space experiments (e.g.
AMS-01, BESS) and will be measured even more accurately in the
near future (e.g. AMS-02 on the ISS). We have developed a two
dimensional (radius and helio-colatitude) model of GCR propagation
in the Heliosphere\cite{six,VO2010}\! that is a function of
measured values of the Solar Wind velocity in the ecliptic plane
($V_0$) and of the neutral sheet tilt angle ($\alpha$), as well as
of stimated values of the diffusion parameter ($K_{0}$). This
model is including curvature, gradient and current sheet drifts,
which are depending on the charge sign of particles and magnetic
field polarity\cite{art_midrift}\!. Particles modulation strictly
depends on the Local Interstellar Spectrum (LIS), up to now not
measured and supposed to be constant with time outside of the
heliosphere. In the model we include also the effects of the
evolving solar activity conditions experienced by a CR particle
inside the heliosphere, due to the time spent by the solar wind to
reach the outer border of the solar cavity. Here we deal with the
differential intensity observed at 1 AU: we do not take into
account the effects of the Earth magnetosphere\cite{mi_jgr}\!.

\section{Stochastic 2D Monte Carlo}\label{sec:2dmodel}

GCR transport in the Heliosphere is described by the Parker
equation\cite{7,8}\!:

\begin{eqnarray}
\frac{\partial U}{\partial t} = & \frac{\partial}{\partial
x_i}\left( K^S_{ij}\frac{\partial U}{\partial x_j} \right)
-\frac{\partial}{\partial x_i}(V_{sw_i}U) +
 \frac{1}{3}\frac{\partial V_{sw_i}}{\partial
x_i}\frac{\partial}{\partial T}(\varrho T U)
-\frac{\partial}{\partial x_i}(\langle v_{d_i} \rangle U)
\label{eq_parker}
\end{eqnarray}

\noindent where $U$ is the CR number density per unit interval of
particle kinetic energy with time ($t$), $T$ is the kinetic energy
(per nucleon), $V_{sw_i}$ the solar wind velocity along the axis
$x_i$, $\langle v_{d_i} \rangle$ is the drift
velocity\cite{art_ste}\!, $K^S_{ij}$ is the symmetric part of the
diffusion tensor and $\varrho$=(T+$2T_{0})/(T$+$T_{0})$, where
$T_{0}$ is particle's rest energy.

As demonstrated by Ito\cite{ito}\!, eq.~\ref{eq_parker} is
equivalent to a set of ordinary stochastic differential equations
that could be integrated with Monte Carlo techniques. The
integration time step ($\Delta t$) is related to the accuracy of
the integrating process and is taken to be proportional to $r^{2}$
($r$ is the distance from the sun) to save CPU
time\cite{Alanko_usoskin2007}\!. We considered\cite{ecrs08}\! the
2D (radius and colatitude) approximation of eq.~\ref{eq_parker}.
The drift velocity is split\cite{art_ste}\! in regular drift
($v_{Di}$) and neutral sheet drift ($v_{ns}$). The model is
depending on some parameters related to the solar activity which
are fine-tuned comparing simulations with experimental
data\cite{ICRC09}\!.

The perpendicular diffusion coefficient has two components, radial
$K_{\perp r}$, and polar $K_{\perp \theta}$. The parallel
diffusion coefficient\cite{9}\! is evaluated by the relation
$K_{||}$=$K_{0}\beta K_{P}(P)(B_{\oplus}/3B)$. Moreover
$(K_{\perp})_{0}$ represents the ratio between perpendicular and
parallel diffusion coefficients: $K_{\perp r}$=$(K_{\perp})_{0}
K_{||}$. Here $K_{0} = 1-6 \times 10^{22}$
cm$^{2}$s$^{-1}$GV$^{-1}$, $\beta$ is the particle velocity, $P$
is the CR particle's rigidity, the $K_{P}(P)$ term takes into
account the dependence on rigidity (in GV), $B_{\oplus} \sim 5$ nT
is the value of heliospheric magnetic field at the Earth orbit,
and $B$ is the magnitude of the Parker field\cite{10}\!. $K_{0}$ is estimated following the procedure discussed in Section 2.2 of\cite{Bobik_2011}\!. The
perpendicular diffusion coefficient in the polar direction
$K_{\perp \theta}$ has been enhanced with respect to $K_{\perp r}$
to reproduce the correct magnitude and rigidity dependence of the
latitudinal cosmic ray proton and electron gradients
\cite{potgieter97}\!.

We use as Heliospheric Magnetic Field (HMF) the Parker field
model\cite{two}\!. The Parker field has been modified\cite{jk89}\!
introducing a small latitudinal component. This modification
increases the magnitude of the HMF in the polar regions without
modifying the topology of the field. The main effect is a lower CR
penetration along polar field lines in the inner part of the
heliosphere, caused by a lower magnetic drift velocity in this
region, as expected from measured data. According to Ulysses data, for periods of low solar activity, we
use a latitudinal dependence\cite{11}\! of the solar wind velocity
on the solar colatitude. Solar wind speed values range from
$V_{min} \simeq$ 400 km/s in the ecliptic plane, to $V_{max}
\simeq$ 750 km/s in the polar regions. Drift effects are included
through analytical effective drift velocities: in the Parker
spiral field we evaluated the three components of drift (gradient,
curvature and neutral sheet\cite{12}\!) that modify the
integration path inside the heliosphere. We adopted the approach
of Potgieter \& Moraal\cite{art_ste}\!, because, in comparison
with other wavy neutral sheet models, it is able to reproduce the
effects in both quite and active solar periods.

\section{Parameters and experimental constraint}\label{sec:proton}

\subsection{Proton LIS}\label{subsec:lis}

One of the inputs of our code is the proton LIS. In order to
estimate the reliability of LIS available in literature we have
considered the proton spectra measured in several periods by
several experiments for rigidity greater than 20 GeV. In this
region we can assume negligible the effects of solar modulation.
Therefore the spectrum, which can be represented by a power law
[$N(R)_{R \geq 20 GV} = \Phi_0 (R/R_0)^{\gamma}$] with a spectral
index $\gamma \simeq -2.7$, can be also considered the LIS.

\begin{figure}[hbt!]
\centering
\includegraphics[width=4.5in]{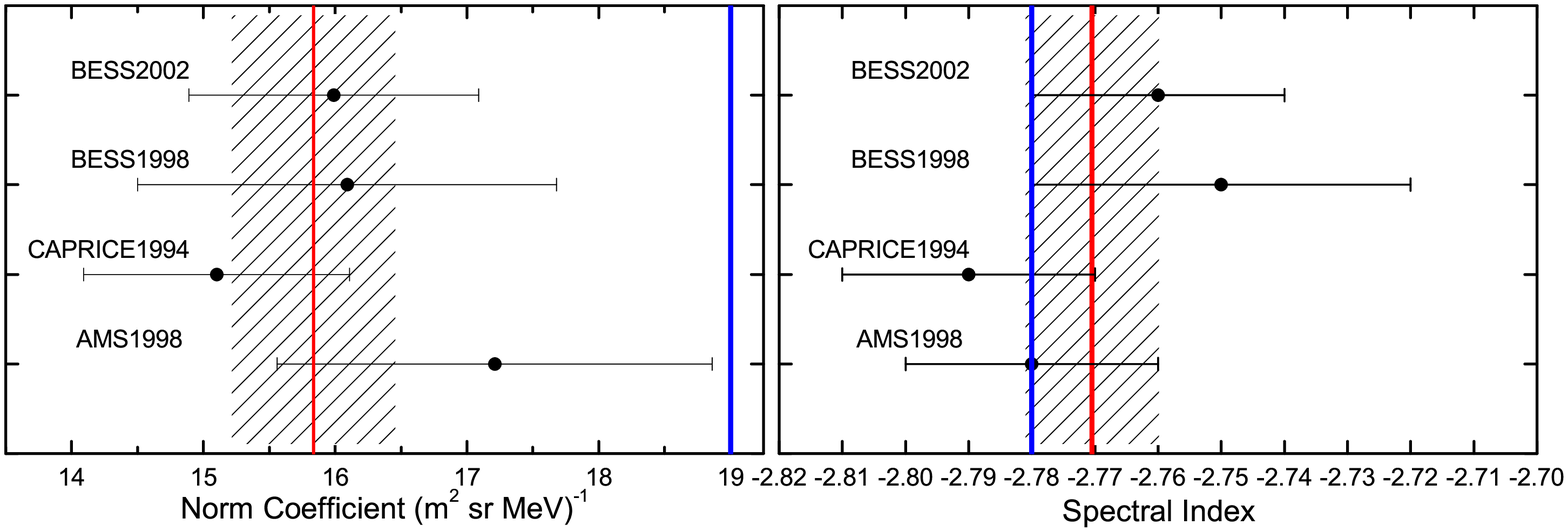}
\caption{Fit (red line) of high energy proton spectrum measured by
several experiments (black dots) and comparison with Burger \&
Potgieter\cite{10}\! LIS (blue line). The spectral index is
consistent with data, but the normalization constant is too
large.} \label{fig_simlis}
\end{figure}

We
focused our analysis on four sets of data: CAPRICE-1994\cite{caprice}\!,
BESS-1998\cite{bess98}\!, AMS-1998\cite{ams01}\! and
BESS-2002\cite{bess02}\!. We estimated the spectral index
($\gamma$) and normalization constant ($\Phi_0$) for each
experiment, then we evaluated an error-weighted average on these
results. We obtained: $\gamma = -2.77 \pm 0.01$ and $\Phi_0 = 15.8
\pm 0.6$ (m$^2$ sr MV)$^{-1}$. In figure \ref{fig_simlis} we
compare our results with the model by Burger \&
Potgieter\cite{10}\!, where it seems to slightly overestimate the
experimental data. We systematically repeated the above analysis fitting all together
the data sets obtaining: $\gamma = -2.77 \pm 0.04$ and $\Phi_0 =
17 \pm 3$ (m$^2$ sr MV)$^{-1}$. Therefore, in the following
calculations we adopt the model by Burger \& Potgieter, corrected
by a scale factor in order to fit experimental data.

\subsection{Latitudinal Gradient}\label{subsec:latgrad}

\begin{figure}[hbt!]
\centering
\includegraphics[width=2.in]{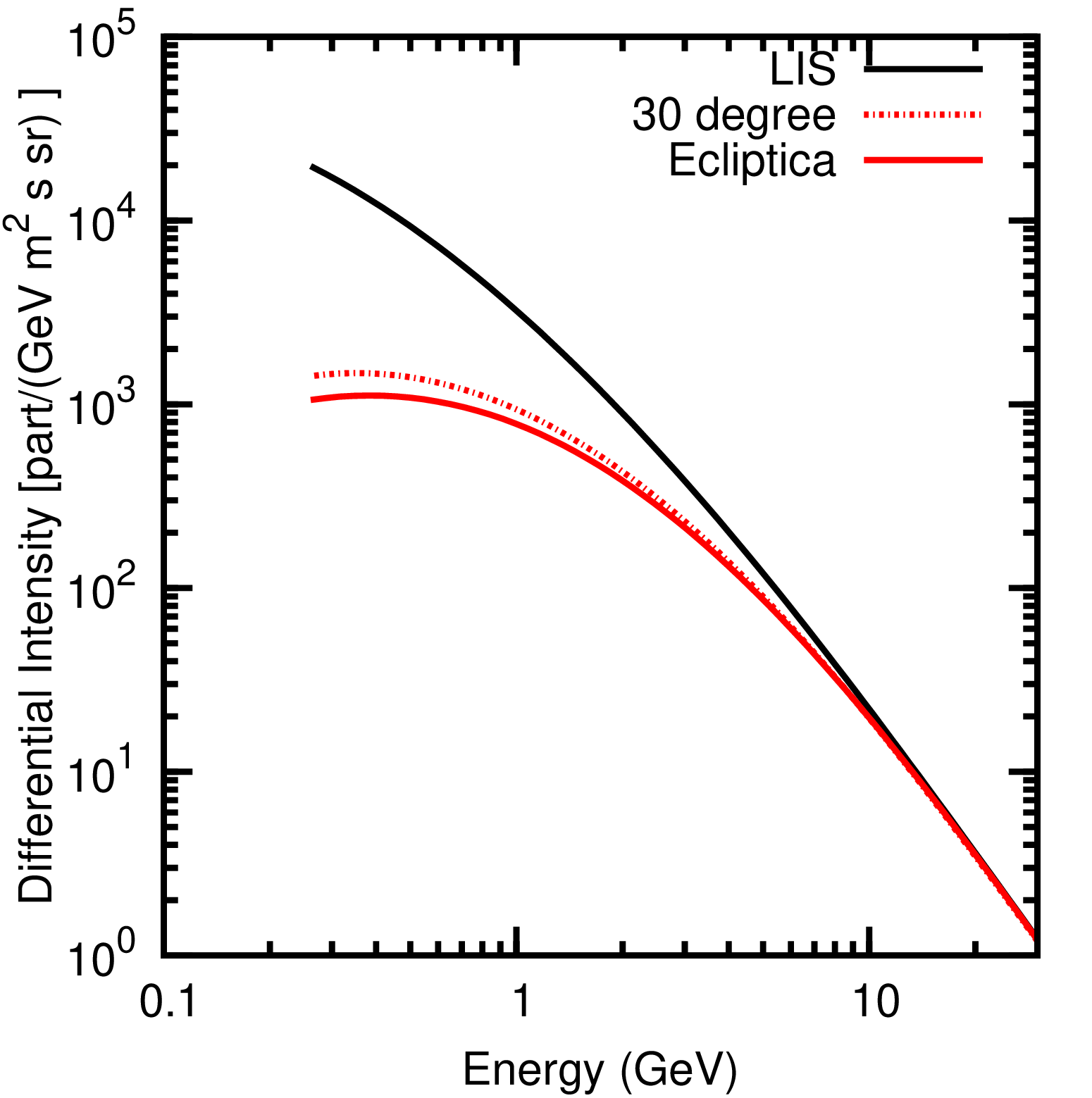}
\includegraphics[width=2.in]{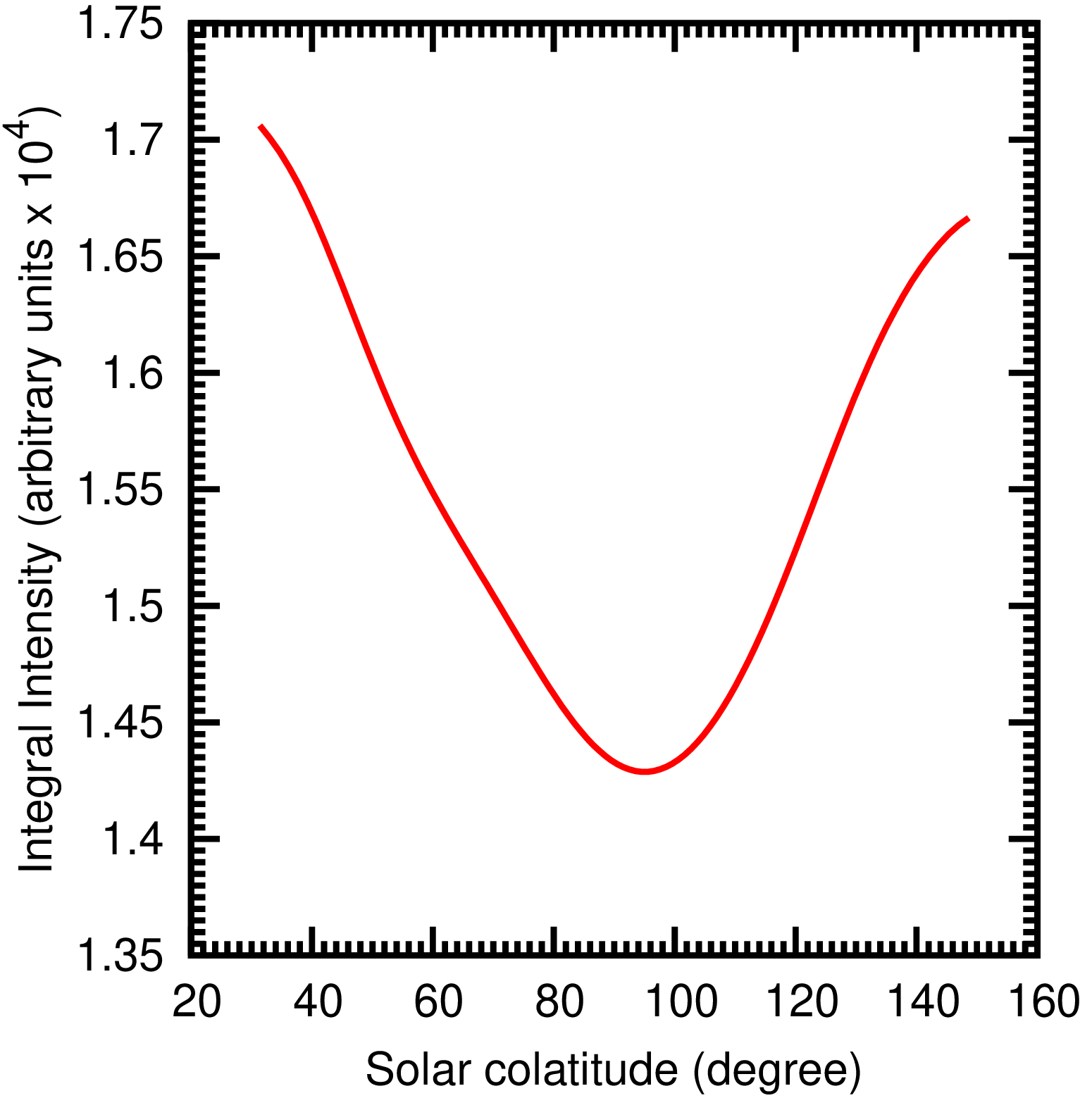}
\caption{Left panel: simulation of the proton differential
intensity on the ecliptic plane is compared with the same quantity
at 30 deg from the poles. Right panel: the latitudinal gradient,
i.e. the integral intensity for $E \geq 200$ MeV vs the solar
colatitude. Simulations have been done using solar parameters
occurred in June 1998.} \label{fig_simlat}
\end{figure}

Observations made by the Ulysses spacecraft in the inner
heliosphere have shown that the latitudinal dependence of CR
protons is significantly less than predicted by classical drift
models\cite{Ulysses}\!. Our model, due to the modification of the
$K_{\perp \theta}$\cite{Bobik_2011}\! parameter in the polar
regions can reproduce the gradient observed by Ulysses (see Fig. 3
in Heber et al.\cite{Ulysses}\!) between the poles and the
ecliptic plane. We find, for the period of AMS-01 data taking
(June 1998), a difference of $\sim$ 16 \% between the ecliptic
plane and a colatitude of $\sim 30^{\circ}$ from the poles. In
Figure \ref{fig_simlat} a simulation of the proton differential
intensity on the ecliptic plane is compared with the one at 30 deg
from the poles (left panel). The latitudinal gradient of the
integral intensity is shown in the right panel of Figure
\ref{fig_simlat}.

\section{Results}\label{sec:res}

\subsection{Proton differential intensity}\label{sec:pres}

\begin{figure}[hbt!]
\centering
\includegraphics[width=2.0in]{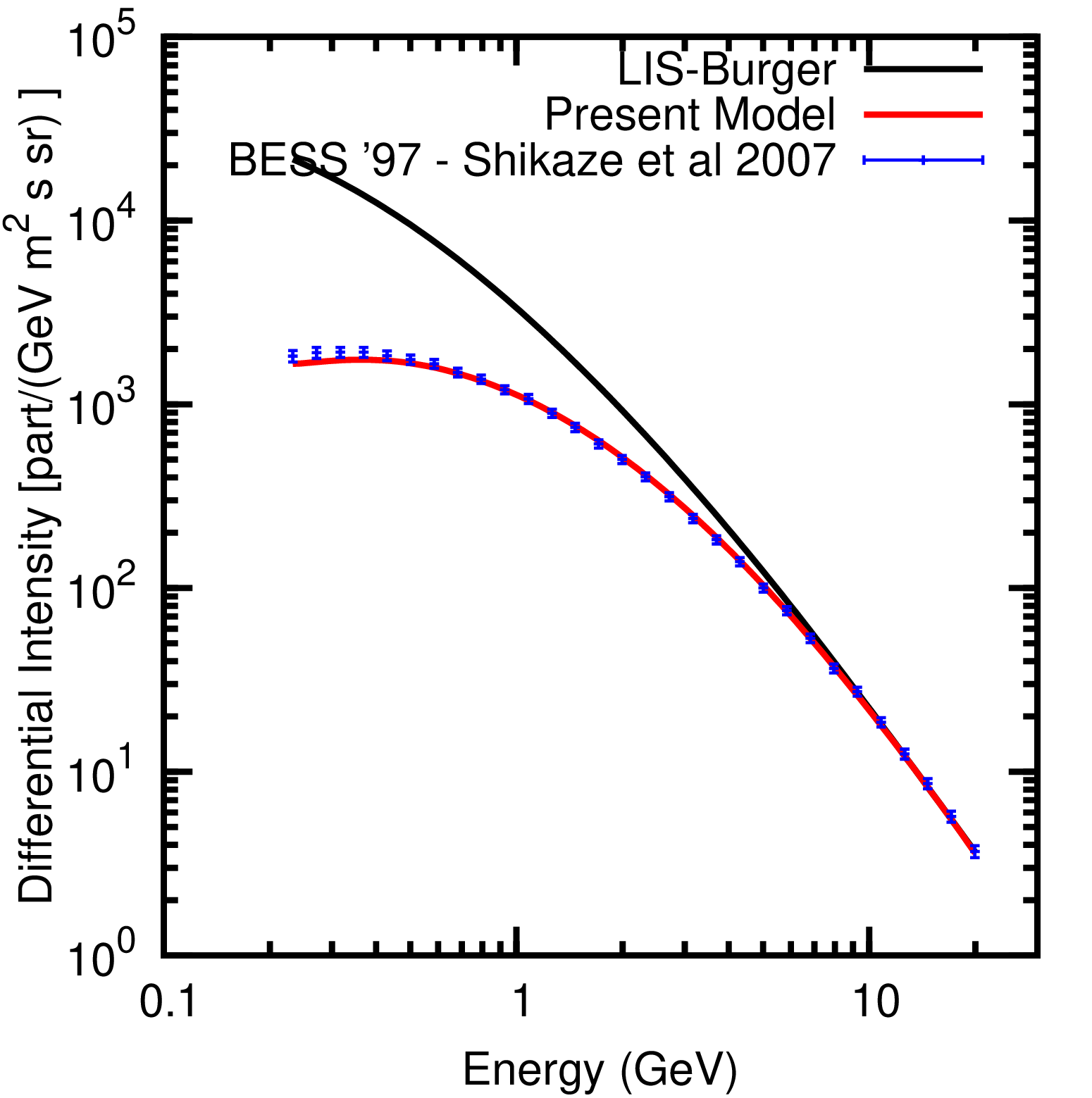}
\includegraphics[width=2.0in]{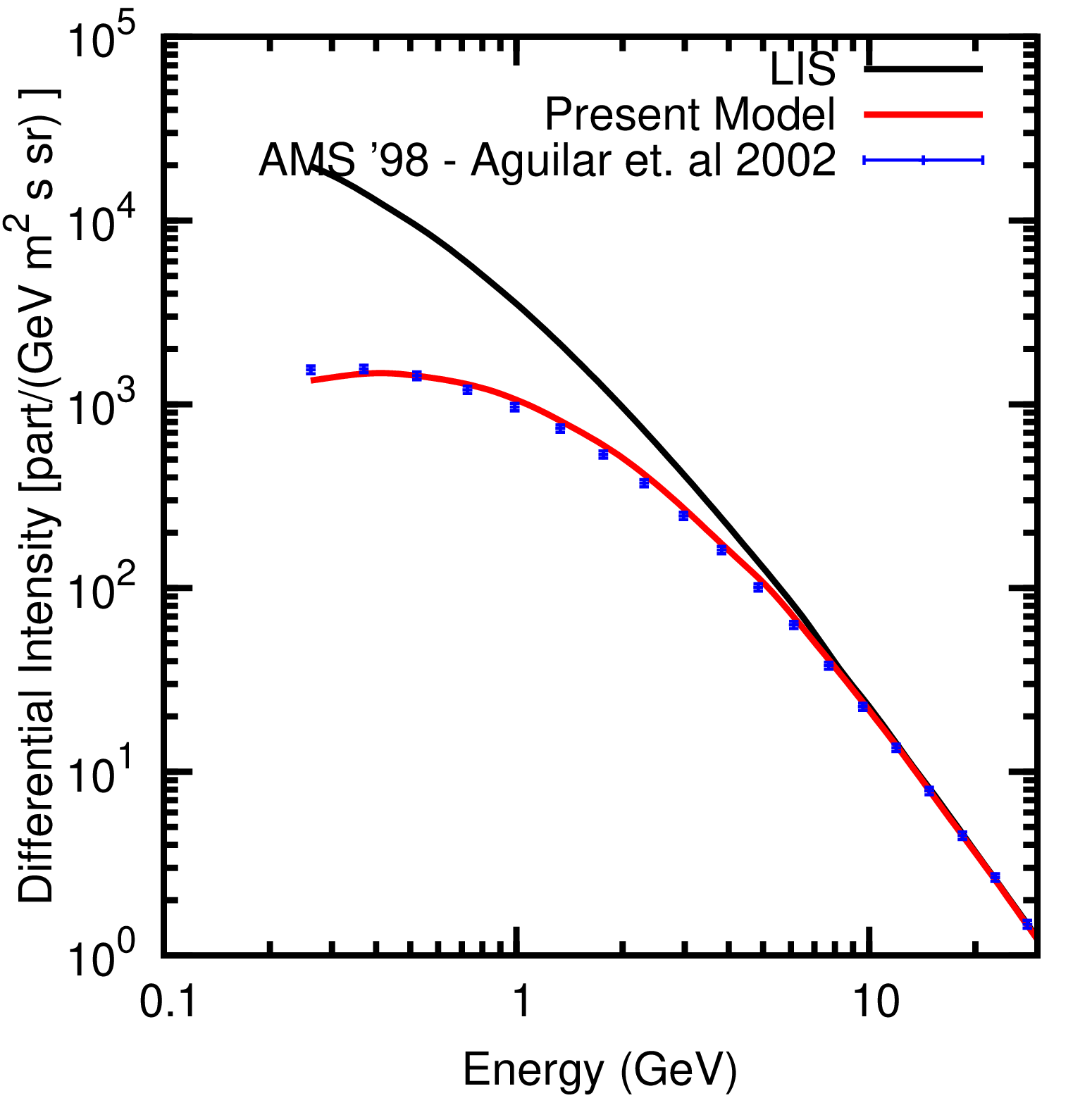}
\includegraphics[width=2.0in]{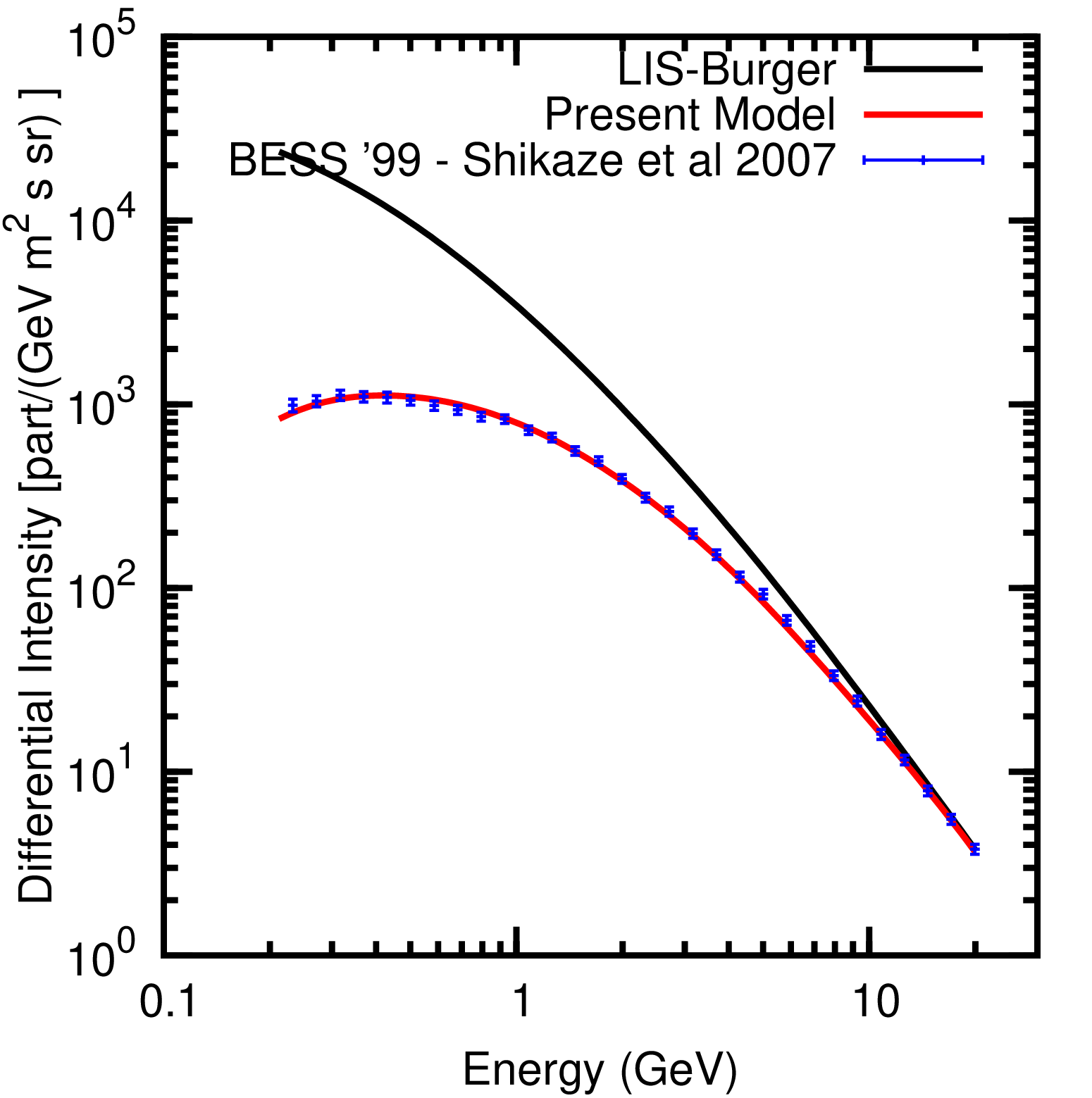}
\includegraphics[width=2.0in]{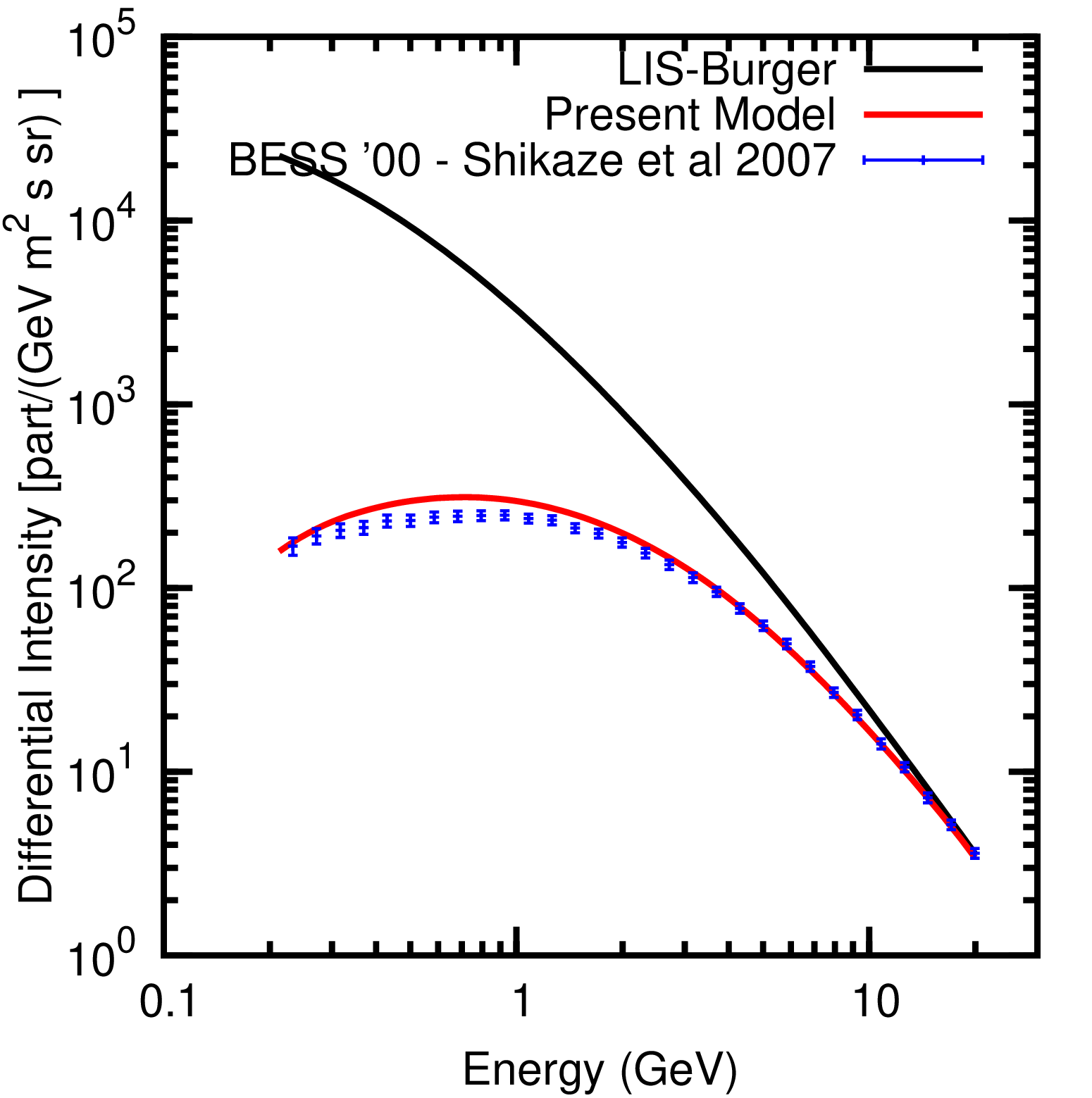}
\includegraphics[width=2.0in]{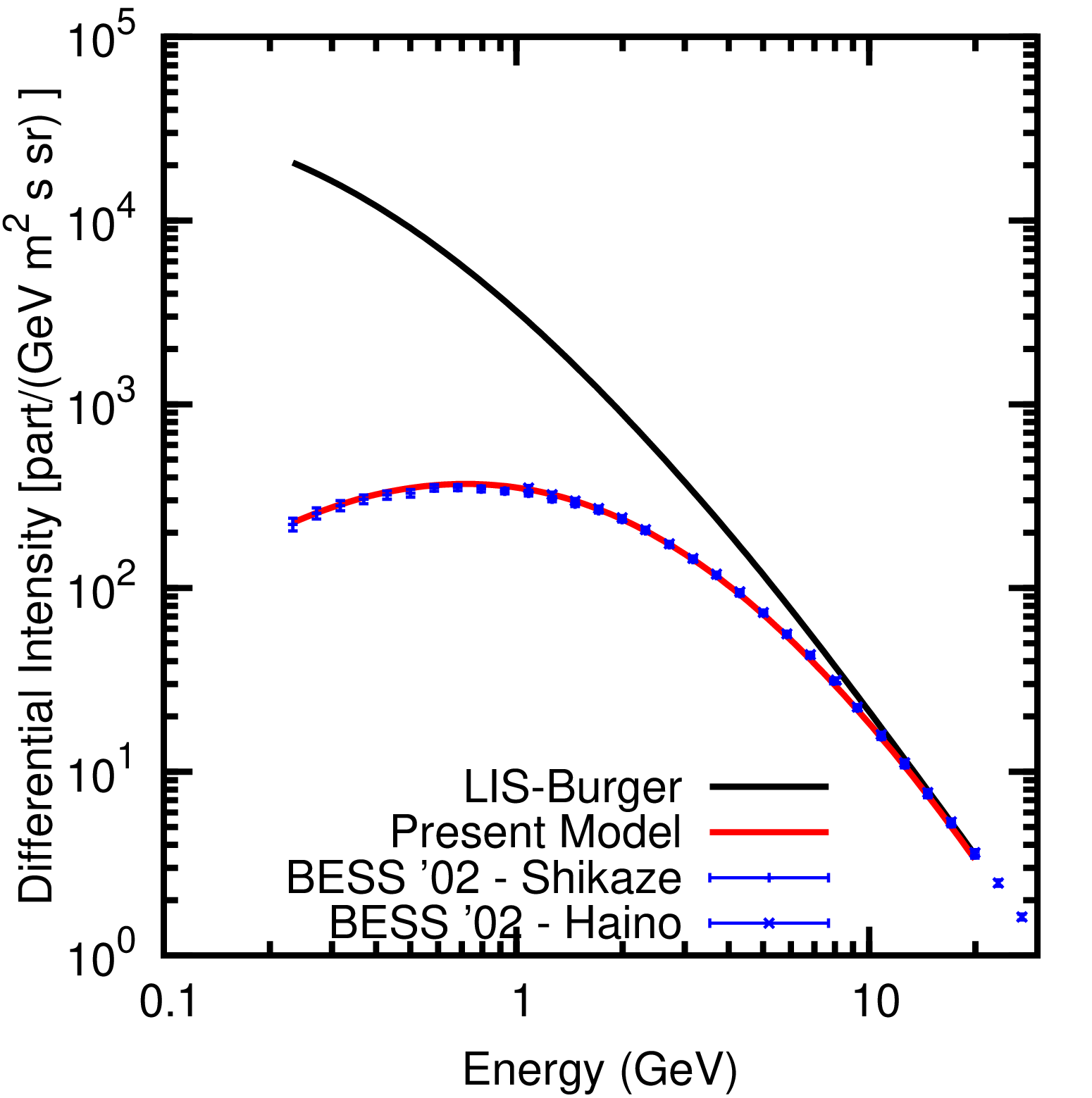}
\includegraphics[width=2.0in]{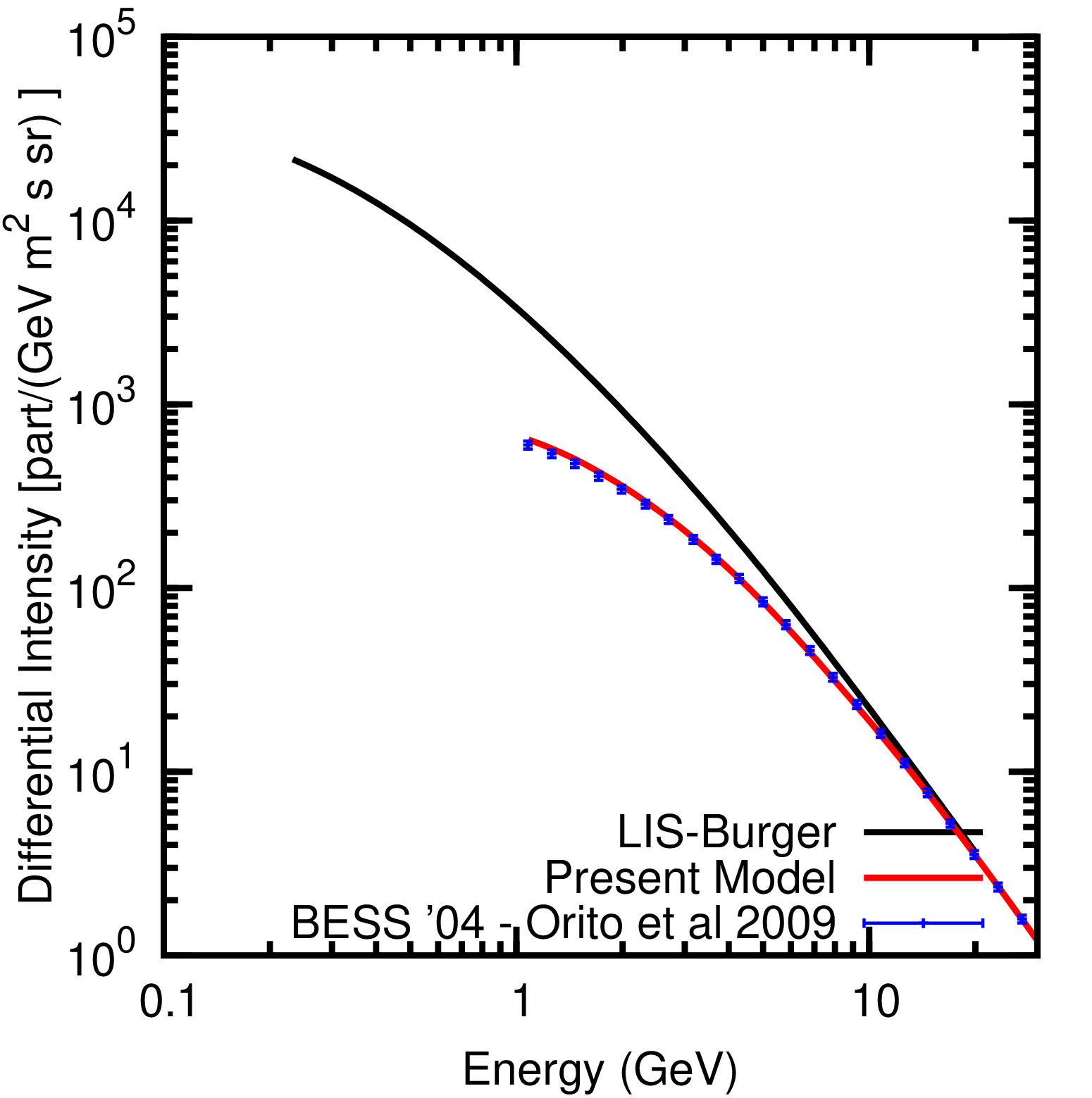}
\caption{Comparison of simulated galactic proton differential
intensity at 1 AU and experimental data: AMS-01\cite{ams01}\!
(1998) and BESS\cite{bess98,bess04}\! (1997, 1999, 2000, 2002,
2004).} \label{fig_sim2}
\end{figure}

We performed the simulations using dynamic values of $K_{0}$,
$\alpha$ and $V_{sw}$: we consider present time parameters in the
inner shell of the heliosphere, but we use the values assumed by
the parameters up to 14 months back in the past at the heliopause.
Results are shown in \mbox{Figure \ref{fig_sim2}}. Simulated
differential intensity with dynamic values shows a very good
agreement with measured data, within the quoted error bars. This
happens in periods with low solar activity and A$>$0, in
comparison with AMS-01\cite{ams01}\!, BESS-97\cite{bess98}\! and
BESS-99\cite{bess98}\!; in periods with high solar activity, in
comparison with BESS-00\cite{bess98}\!; and in periods with a
lower solar activity and A$<$0, in comparison with
BESS-02\cite{bess02}\! and BESS-04\cite{bess04}\!. This means that
our description of the Heliosphere improves the understanding of
the complex processes occurring inside the solar cavity.

\subsection{AMS-02 Predictions}\label{sec:ams02}

\begin{figure}[hbt!]
\centering
\includegraphics[width=2.50in]{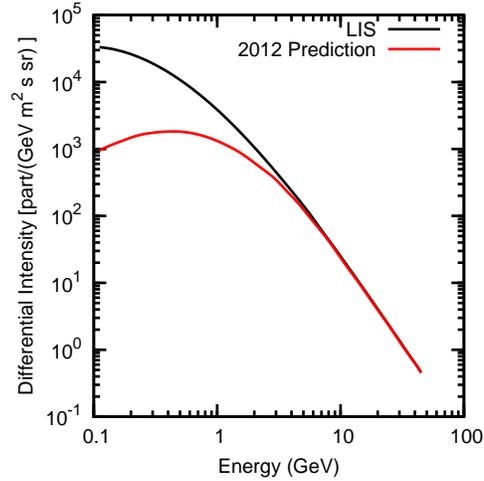}
\caption{Prediction of modulated proton differential intensity as
will be measured by AMS-02 experiment in January 2012.}
\label{fig_sim3}
\end{figure}

Our simulation code has been used to predict CR differential
intensity for future measurements. The assumption is that
diffusion parameter, tilt angle and solar wind speed show a
near-regular and almost periodic trend. The periodicity occurs
after two consecutive 11-years solar cycles. Using SIDAC (Solar
Influences Data Analysis Center) data we selected periods with a
nearly similar solar activity conditions and same solar field
polarity of the simulation time: therefore approximately 22 years
in advance. We concentrate our simulations on the AMS-02 mission,
that will be installed on the ISS in February 2011, for a period
approaching the solar maximum: January 2012. Results are shown in
Fig. \ref{fig_sim3}.

The Sun is currently in an unpredicted long duration solar
minimum, that forced scientists to review all their estimations
for the next solar cycle, therefore this prediction could be
object of revision in the next future. AMS-02 is expected to
collect, in a few years of operation, more than $10^{10}$ protons
with energy $\geq$ 1 GeV, and $\sim 10^{6}$ with energy $\geq$ 1
TeV.

\section{Conclusions}\label{sec:conc}

We built a 2D stochastic Monte Carlo code for particles
propagation inside the heliosphere. Our model takes into account
drift effects and shows a good agreement with measured values, in
periods with positive as well as with negative polarity. Proton
spectra, as predicted by the model, are decreasing with increasing
tilt angle and solar wind velocity. We use as input LIS the model
published by Burger \& Potgieter\cite{10}\!, corrected in order to
fit the high energy measured intensity. Recent measurements have
pointed out the needs to reach a high level of accuracy in the
modulation of the differential intensity, in relation to the
charge sign of the particles and the solar field
polarity\cite{mi_ap}\!. This aspect will be even more crucial in
the next generation of experiments.

\end{document}